\documentclass{PoS}
\usepackage{amsmath}
\newcommand{\cwr}[1]{c^{Wr}_{#1}}

\title{Chiral expansion for $\pi^0$ decays}
\ShortTitle{Chiral expansion for $\pi^0$ decays}

\author{\speaker{Karol Kampf}\thanks{I would like to thank the organizers for creating a pleasant atmosphere during the conference. I am grateful to  B.~Moussallam, J.~Novotn\'y and R.~Rosenfelder for their interest in the work and valuable comments to the text. This work is
supported in part by the European commission MRTN FLAVIAnet [MRTN-CT-2006035482], Center for Particle Physics [LC 527] and
GACR [202/07/P249].}\\
         Paul Scherrer Institut, CH-5232 Villigen PSI, Switzerland\\
         Charles University, Faculty of Mathematics and Physics, V Hole\v{s}ovi\v{c}k\'ach 2, Prague, Czech Rep.\\
        E-mail: \email{karol.kampf@psi.ch}}
\abstract{New ongoing experimental activities that have direct reference to $\pi^0$ decay modes call for a new theoretical study in this area.
We will summarize some details and interesting facts that concern main decay modes of this lightest meson.}

\FullConference{International Workshop on Effective Field Theories: from the pion to the upsilon \\
		February 2-6 2009\\
		Valencia, Spain}

\renewcommand{\logo}
{\parbox{\textwidth}{\begin{flushright}
PSI-PR-09-05
\end{flushright}\vspace*{-0.3cm}
\includegraphics{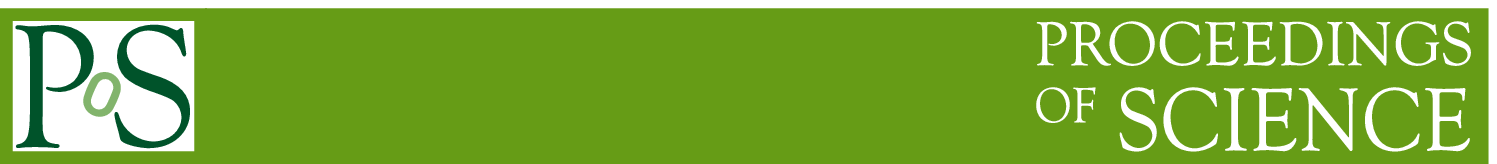}}
}

\begin{document}

\section{Introduction}

The subject of this conference ``from the pion to the upsilon'' covers a wide range of hadron physics. The $\pi^0$ meson has a prominent position among all these particles as being the lightest state of them.  Its primary decay mode is thus $\pi^0 \to \gamma\gamma$ which is connected with the famous Adler-Bell-Jackiw triangle anomaly \cite{ABJ}. There are new ongoing experimental efforts which can potentially, among other things, tell us some new or more accurate information about $\pi^0$. This potential can be found for example in the following experiments: Dirac, KTeV, PrimEx and NA48/2.

This is the main motivation for us to study the present discrepancy between the experiment and theoretical predictions. We will focus on four most important allowed decay modes of $\pi^0$: $\gamma\gamma$, $e^+e^-\gamma$, $e^+e^-e^+e^-$, $e^+e^-$ (with branching ratios \cite{pdg}: $0.98798(32)$, $0.01198(32)$, $3.14(30)\times 10^{-5}$, $6.46(33)\times 10^{-8}$, respectively). For this purpose one can use two-flavour chiral perturbation theory (ChPT, for a review see \cite{bijnenshere}) which can simply incorporate corrections to the current algebra result attributed either to $m_{u,d}$ masses or electromagnetic corrections with other effects hidden in the low energy constants (LECs), denoted by $c_i^W$ at next-to-leading order (NLO).
However, phenomenologically richer $SU(3)$ ChPT must be also employed in order to obtain numerical prediction. This is especially true for the studied anomalous processes as in this case the initial symmetry for the two flavour case must be extended and the number of monomials in $SU(2)$ increases \cite{bijnens01}.

Let us stress that we limit our focus in this article on ``standard on-shell'' decays. It is clear that both on-shell and off-shell or semi-of-shell vertices, especially $\pi^{0(*)} \gamma^*\gamma^{(*)}$, play a crucial role in many other experiments, from the famous $g-2$ (cf. \cite{g2}) via virtual photons stemming from $e^+e^-$ (see e.g. the recent paper \cite{rosner})
to astrophysics. Our first aim is the common formulation of these interrelated processes in the given formalism at the given order (either NLO or NNLO) motivated by the precision of the present or near-future experiments. This can be also viewed in the more ambitious perspective of the fundamental physics searches in the low-energy physics and it thus represents a complement to existing efforts in this direction (as is for example the low energy physics study at the high-intensity proton facility at PSI).

\section{$\pi^0 \to \gamma \gamma$}

As stated in the introduction the $\pi^0 \to \gamma\gamma$ is a crucial decay mode of $\pi^0$. It saturates its decay width with almost 99\% and plays an important role in the further decay modes (see the following sections). The history of $\pi^0\to\gamma\gamma$ is going back to Steinberger's calculation \cite{steinberger} (this year the 60th anniversary!). This calculation and its connection with the famous Adler-Bell-Jackiw anomaly is now part of almost every modern textbook on quantum field theory, not necessary focusing on QCD (see e.g. \cite{georgi}, cf. also \cite{horejsi}). The prediction estimated from the chiral anomaly using current algebra agrees surprisingly very well with experiment. A first attempt to explain the small existing deviation from the measurement was made by Y.~Kitazawa \cite{kitazawa}. The experimental situation then was the same as it is now according to the accepted numbers by the particle data group (PDG) \cite{pdg}. At that time a new experimental prediction from CERN-NA030 \cite{exp:cern} suggested a smaller value for the partial width $7.25\pm 0.23$ eV (statistical and systematic errors combined in quadrature). Older experiments (Tomsk, Desy and Cornell \cite{exp:prim}), seemed not to be so precise ($7.23\pm0.55$,\ $11.7\pm1.2$,\ $7.92\pm0.42$ eV, respectively); they relied on the so-called Primakoff effect \cite{primakoff} that is based on measuring the cross section for the photoproduction of the meson in the Coulomb field. The more precise number from the direct measurement at CERN motivated Y.~Kitazawa to explain the $8\sim 9 \%$ discrepancy by including QED correction and the $\eta/\eta'$ contribution. These corrections were not, however, large enough to explain the discrepancy which was attributed by the author to a possible $\pi(1300)$ contribution. Furthermore, it was found out in this work that the contribution from multi-pion states must be small. This was verified explicitly also within ChPT with the remarkable observation \cite{donbij} that at one-loop order there are no chiral logarithms (either from pions or kaons). The $\pi$-$\eta$-$\eta'$ mixing and electromagnetic correction were reconsidered relatively recently in \cite{reco}.

The spread in the data basis of the PDG, summarized in the previous paragraph, shows, however, that the quoted errors seem to be underestimated \cite{bernstein}. The present situation fortunately looks more optimistic as the world average accuracy of 8~\% is planned to be improved to the level of one or two percents in ongoing experiment PrimEx at JLab \cite{bernstein}. This was the main motivation for a new study of $\pi^0\to\gamma\gamma$ in \cite{KM}. The correction to the chiral anomaly due to the finite mass of light quarks was reconsidered using strict two-flavour ChPT at NNLO. We will summarize here this remarkably simple result (note that it involves a two-loop calculation and that it represents formally a full $O(p^8)$ result). Defining a reduced $T$ amplitude
\begin{equation}
A = e^2 \varepsilon_{\mu\nu\alpha\beta} \epsilon_1^{*\mu} \epsilon_2^{*\nu} k_1^\mu k_2^\nu\ T \,,
\end{equation}
we have for the partial decay width
\begin{equation}
\Gamma_{\gamma\gamma} = \frac{\pi}{4} \alpha^2 m_{\pi^0}^3 |T|^2\,.
\end{equation}
Up to and including next-to-next-to-leading order corrections
\begin{align}
F_\pi T_{NNLO} &=  {1\over 4\pi^2}
+{16\over3}{m_\pi^2}\left(-4\cwr{3}-4\cwr{7}+\cwr{11}\right)
+{64\over9}{B(m_d-m_u)}(5\cwr{3} + \cwr{7}+2\cwr{8})
\notag\\%%%%%%%%%%%%%%%
&+ {M^4\over 16\pi^2 F^4}\,L_\pi
\left[ {3\over256\pi^4} +
{32F^2\over3}\left(2\cwr{2}+4\cwr{3}+2\cwr{6}+4\cwr{7}-\cwr{11}\right)
\right]
\notag\\%%%%%%%%%%%%%%%
&+{32 M^2 B(m_d-m_u)\over 48\pi^2 F^4}\,L_\pi\,
\left[-6\cwr{2}-11\cwr{3}+6\cwr{4}-12\cwr{5}-\cwr{7}-2\cwr{8}\right]
\notag\\%%%%%%%%%%%%%%%
&-{M^4\over24\pi^2 F^4}\,\left( {1\over16\pi^2}L_\pi \right)^2
+{M^4\over F^4} \lambda_+   +{M^2 B(m_d-m_u) \over F^4} \lambda_-
+{B^2(m_d-m_u)^2\over F^4} \lambda_{--}\ ,
\end{align}
where the chiral logarithm is denoted by $L_\pi=\log{m_\pi^2\over\mu^2}$
and $\lambda_+$, $\lambda_-$, $\lambda_{--}$
can be expressed as follows in terms of renormalized
chiral coupling constants ($d^{Wr}$ refer to combinations of couplings from
the NNLO Lagrangian, i.e. of order $p^8$ in the anomalous sector),
\begin{eqnarray}
&& \lambda_+ = {1\over\pi^2}\left[
-{2\over 3} d_+^{Wr}(\mu) -8c_6^r  -{1\over4}(l_4^r)^2
+{1\over 512\pi^4} \left( -{983\over288} - {4\over3}\zeta(3)
+3\sqrt3\, {\rm Cl}_2(\pi/3)
\right)\right]
\nonumber\\
&& \phantom{\lambda_+ = } +{16\over3} F^2
\left[\,8l_3^r(\cwr{3}+\cwr{7})+l_4^r(-4\cwr{3}-4\cwr{7}+\cwr{11})\right]
\nonumber\\
&& \lambda_-= {64\over9}\left[
d_-^{Wr}(\mu) +F^2 l_4^r\,(5\cwr{3}+\cwr{7}+2\cwr{8}) \right]
\nonumber\\
&& \lambda_{--}= d_{--}^{Wr}(\mu)-128F^2 l_7 (\cwr{3}+\cwr{7})\ .
\end{eqnarray}

All effects that were carefully studied e.g. in \cite{kitazawa} and \cite{reco} are now hidden in the LECs and chiral logarithms $L_\pi$ (with the exception of QED corrections that must be added by hand to the latter formula, see \cite{KM}). This inevitably leads us to the phenomenological study of every piece in the last formula. Intuitively, one expects that a three flavour input can give us some new information. We know that there exist at least data on $\eta$ and $\eta'$, but on top of that, as discussed also in the introduction, in the two-flavour case the number of relevant NLO LECs ($\cwr{i}$) is bigger than the number of three-flavour $C^{W}_i$s. The advantages of $SU(3)$ can be included consistently via a modified three-flavour counting, where $m_{u,d}$ count as $O(p^2)$ and $m_s\sim O(p)$ (for details see \cite{KM}).

It is interesting to realize that the absence of chiral logs at NLO and their smallness at NNLO signals a fast convergence of the ChPT series. Taking the chiral logs as an estimate of the size of chiral corrections one can see the importance of $F_\pi$ in $\pi^0\to\gamma\gamma$ decay. The $F_\pi$, on the other hand, is determined from the weak decay of $\pi^+$ based on the standard $V-A$ interaction. The new proposed variant of this interaction assumes contributions of right-handed current which would lead to a change of $F_\pi$ \cite{bernard}. Determination of this constant directly from $\pi^0$ lifetime can provide constraints on such contributions. Our best estimate leads to (n.b. $F_\pi = 92.22$ MeV)
\begin{equation}
\Gamma_{\gamma\gamma}= (8.09 \pm 0.11)\ {\rm eV }.
\end{equation}

\section{$\pi^0 \to e^+e^- \gamma$}
The internal conversion of one of photons in $\pi^0\gamma\gamma$ into $e^+e^-$, a so-called ``Dalitz pair'' \cite{dalitz} leads us to the second decay mode with a branching ratio $\sim 1.198 \pm 0.032\%$. Knowing relatively precisely this branching ratio (with the same absolute error as for the $\pi^0\to\gamma\gamma$ mode) means that one could in principle use also the Dalitz decay to extrapolate the total decay width. This can serve as an independent possibility\footnote{However, having the same experimental precision as for $\pi^0\to\gamma\gamma$, this decay mode has the disadvantage that the error of the life-time is larger approximately by a factor of two.} how to measure the life-time of $\pi^0$. What is however interesting in connection with $\pi^0\to e^+ e^- \gamma$ is the differential decay width. Even the integrated $\Gamma_{e^+e^-\gamma}$ is small and thus the study of QED and chiral corrections seems not to be needed. Indeed, the QED correction is a tiny number \cite{lautrup71}
\begin{equation}\label{tinynumber}
\frac{\Gamma^\text{QED}_{e^+ e^-\gamma}}{\Gamma_{\gamma\gamma}}=
\Bigl(\frac{\alpha}{\pi}\Bigr)^2
\Bigl(\frac{8}{9}\ln^2{\frac{M_{\pi^0}}{m}}-\frac{19}{9}\ln{\frac{M_{\pi^0}}{m}}
+2\zeta(3) +\frac{137}{81} -\frac{2\pi^2}{27} +
O(\frac{m}{M_\pi})\Bigr)= 1.04\times 10^{-4}\,,
\end{equation}
if compared with LO \cite{dalitz}:
\begin{equation}
\frac{\Gamma^\text{LO}_{e^+ e^-\gamma}}{\Gamma_{\gamma\gamma}}= \frac{\alpha}{\pi}
\Bigl(\frac43 \ln \frac{M_{\pi^0}}{m} - \frac73 + O(\frac{m^2}{M_\pi^2}) \Bigr) =
0.01185\,.
\end{equation}
However, it turns out that the corrections to the differential decay are indeed important. The reason is that there is a part of the phase space where, roughly-speaking, the correction to the differential decay width is positive and a part where it is negative; and only summing these parts together gives us the small number in~(\ref{tinynumber}). It is clear now, that in physically relevant applications, when we have to cut some parts of the phase space, these corrections can become important.

A detailed study of the Dalitz decay can be found in \cite{KKN}. It extended Mikaelian and Smith's calculation of the QED corrections \cite{mikaelian72} and by this it clarified doubts about the applicability of Low's theorem \cite{tpe2}.
The work \cite{KKN} thus provides a detailed analysis of NLO radiative corrections to the Dalitz decay amplitude.
We have included there the off-shell pion-photon transition form factor which requires a treatment of non perturbative strong interaction effects.
The one-photon irreducible contributions, which had been neglected in \cite{mikaelian72}, were also calculated.
The relevance of these corrections was demonstrated for the slope parameter
$a_\pi$ of the pion-photon transition form factor. Our prediction for $a_\pi=0.029 \pm 0.005 $
is in good agreement with the determinations
obtained from the (model dependent) extrapolation of
the CELLO and CLEO data.

The usually omitted contribution of the two-photon exchange represents approximately more than $15\ \%$ in the $a_\pi$ prediction. This indicates that a detailed analysis of the radiative corrections is also inevitable in the following decay modes.

\section{$\pi^0 \to e^+ e^- e^+ e^-$}

Due to the Young-Landau theorem \cite{yang} we know that the pion cannot be a $J=1$ state. To verify experimentally whether it is a (pseudo)scalar it is very difficult to use directly $\pi^0 \to \gamma\gamma$ as it is not possible to measure the photon polarization (due to the narrow angle between $e^+$ and $e^-$ emitted from the real photon). It was thus suggested in \cite{krollwada} to use the double-internal conversion, the so-called double-Dalitz decay. The experiment was performed at Nevis Lab \cite{samios} in a bubble chamber with the following result for the branching ratio (today's PDG number):
\begin{equation}
\frac{\Gamma_{e^+e^-e^+e^-}^\text{PDG}}{\Gamma_{tot}} = (3.18\pm 0.30)\times 10^{-5}
\end{equation}
and it confirmed the negative parity of $\pi^0$ known from the previous indirect measurements via the cross-section of $\pi^-$ capture on deuterons. However, the significance of this direct measurement was only 3.6 $\sigma$. Last year the long standing experimental gap was filled with a new measurement in the KTeV-E799 experiment at Fermilab \cite{abouzaid2} giving a branching ratio (including the radiative final states above a certain cut as tacitly assumed for all Dalitz modes)
\begin{equation}
\frac{\Gamma_{e^+e^-e^+e^-}^\text{KTeV}}{\Gamma_{tot}} = (3.46\pm 0.19)\times 10^{-5}\,,
\end{equation}
which is in good agreement with the previous experiment. In addition to the precisely verified parity of $\pi^0$ (which represents its best direct determination) this experiment sets the first limits on the parity and CPT violation for this decay. More precisely, having a $\pi^0\gamma^*\gamma^*$ vertex  $C_{\mu\nu\rho\sigma} F^{\mu\nu}F^{\rho\sigma} \pi^0$ we can study, using the following decomposition (for details see \cite{barker})
$$
C_{\mu\nu\rho\sigma} = \cos\zeta \varepsilon_{\mu\nu\rho\sigma} + \sin\zeta {\rm e}^{i\delta} (g_{\mu\rho}g_{\nu\sigma} - g_{\mu\sigma}g_{\nu\rho})\,,
$$
the parameters $\zeta$ and $\delta$ which represent parity mixing and CPT violation parameters. For details see \cite{abouzaid2}; for example their limit on the mixing assuming CPT conservation is $\zeta<1.9^\circ$.

A detailed analysis of the radiative corrections in \cite{barker} showed that they seem to be very important in extracting physically relevant quantities. This motivates us to reopen this subject \cite{KKN2} in the same manner as was done in \cite{KKN}. The simply looking task of attaching another Dalitz pair on the virtual photon line is complicated (in the defined power-counting) by the necessity to include a pentagonal diagram \cite{barker}. This strengthens the need of a correct description of the off-shell $\pi^0\gamma^*\gamma^*$ vertex, which can be, on the other hand, directly studied in the next mode.

\section{$\pi^0 \to e^+ e^-$}
Last but not least let us briefly mention a decay mode which is directly connected with the fully off-shell $\pi^0 \gamma^* \gamma^*$ vertex and represents the best candidate for studying not yet well-understood effects of QCD or (if one prefers) effects of eventual new physics. This is supported by an existing experiment at Fermilab (KTeV E799-II) \cite{abouzaid1}. Comparing with previous measurements their result has increased significantly the precision and made thus the most important contribution to present PDG's average
$$
\frac{\Gamma_{e^+e^-}^\text{PDG}}{\Gamma_{tot}} = (6.46 \pm 0.33)\times 10^{-8}\,.
$$
Apart from the imaginary part which can be calculated in a model independent way (and set the so-called unitary bound) the real part depends on the chosen model (for a review see \cite{dorokhov07}). It is important that the precision of the KTeV experiment can already distinguish among given models and can be also naturally used as a test for new physics (as was done e.g. in \cite{chang}).

Discussions of the appropriate description of $\pi^0\gamma^*\gamma^*$ would go beyond the scope of this paper. However, setting the limits on this vertex by calculating radiative corrections is of great importance. A new calculation in this direction~\cite{dorokhov08} shows that the radiative corrections used in the extrapolation are indeed under control. This calculation together with the correct description of the Dalitz decay (which was used in normalization in order to improve systematic error) is under investigation \cite{KKN3}.

\section{Summary}
The new experimental activities in the low energy physics that concern directly $\pi^0$ decay modes call for a more detailed theoretical study in this area. We have discussed the main points that concern four most important allowed decay modes of the lightest meson, namely: $\gamma\gamma$, $e^+ e^- \gamma$, $e^+ e^- e^+ e^-$ and $e^+ e^-$. These processes are connected partly already in experiments, in order to minimize systematic uncertainties and also by the theory as all of them rely on the $\pi^0\gamma\gamma$ vertex. A common treatment is thus useful and important in order to understand all phenomena. In our works we have focused on chiral and QED corrections in order to prepare the ground for the discussion of the non-perturbative effects or eventual new physics.

\end{document}